\documentclass[twocolumn,showpacs,english,prb,preprintnumbers,amsmath,amssymb,floatfix]{revtex4}
\usepackage[T1]{fontenc}
\usepackage[latin9]{inputenc}
\usepackage{babel}
\usepackage{graphicx}
\usepackage{epsfig,psfrag}
\usepackage{dcolumn}
\usepackage{bm}
\usepackage{color}
\usepackage{esint}

\begin{document}

\title{Emerging Dirac and Majorana fermions for carbon nanotubes with proximity-induced pairing and spiral magnetic field}

\author{Reinhold Egger}

\affiliation{ Institut f\"ur Theoretische Physik, Heinrich-Heine-Universit\"at, D-40225  D\"usseldorf, Germany }

\author{Karsten Flensberg}

\affiliation{Niels Bohr Institute, University of Copenhagen, Universitetsparken 5, DK-2100 Copenhagen, Denmark}

\date{\today}

\begin{abstract}
We study the low-energy bandstructure of armchair and small-bandgap
semiconducting carbon nanotubes with proximity-induced
superconducting pairing when a spiral magnetic
field creates strong effective spin-orbit interactions
from the Zeeman term and a periodic potential from the orbital part.
We find that gapless Dirac fermions can be generated by variation of
a single parameter. For a small-bandgap semiconducting tube with the field in the same
plane, a non-degenerate zero mode at momentum $k=0$ can be induced,
allowing for the generation of topologically protected Majorana fermion end
states.
\end{abstract}

\maketitle

\section{Introduction}

The electronic properties of single-wall
carbon nanotubes (CNTs) have been studied for almost two decades by now and
are generally thought to be well understood.\cite{cntreview}
The electronic structure of a CNT on energy scales below
$\hbar v_F/R$, with radius $R$ and Fermi velocity
$v_F\simeq 8\times 10^5$~m/s, is captured by the low-energy approach,
where one retains only the lowest transverse momentum bands and disregards
trigonal warping.  We focus on small-bandgap semiconducting
CNTs without primary gap, where the curvature-induced bandgap is $E_g \propto \cos(3\theta)/R^2$
and typically of order meV in experiments. Here, $\theta$ is the chiral angle,\cite{cntreview} and
only armchair tubes, $\cos(3\theta)=0$, are metallic.  Two interesting
developments concern spin-orbit interactions (SOI) and proximity-induced
pairing correlations. First, Coulomb blockade
spectroscopy experiments for ultraclean CNTs\cite{kuemmeth,jespersen} have
confirmed the existence of the theoretically expected but rather small
curvature-induced intrinsic SOI.\cite{ando,egger,paco,weiss}
Although a proposal exists to design tunable SOI
in graphene by deposition of suitable adatoms,\cite{prx}
this idea does not readily apply to CNTs.
Second, proximity-induced superconductivity was
experimentally demonstrated and has been usefully
exploited.\cite{prox1,prox2,prox3,prox4}
However, only few theoretical studies\cite{smitha,sau,loss1}
have addressed the corresponding pairing terms
in the CNT low-energy theory.

\begin{figure}[t]
\centering
\includegraphics[width=8cm]{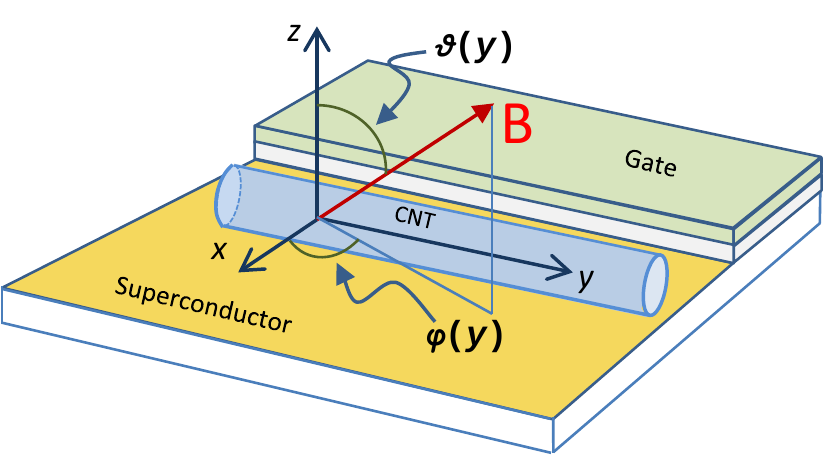}
\caption{\label{fig1} (Color online) Schematic set-up of a carbon
 nanotube proximity-coupled to a superconductor, in 
a spatially dependent magnetic field with
angles $\varphi(y)$ and $\vartheta(y)$, see Eq.~\eqref{field}.
 The chemical potential on the tube is assumed to be tunable 
via a nearby gate electrode.}
\end{figure}

In this work, we demonstrate that the combined effects of strong SOI and
proximity-induced superconductivity in CNTs are responsible for
emergent gapless Dirac fermions and Majorana bound states.
The schematic set-up considered here is shown in Fig.~\ref{fig1}.
Apart from the spiral magnetic field used to generate strong
spin-orbit couplings, the set-up is similar to recent proposals
for semiconducting nanowires.\cite{lutchyn,felix} In both cases, despite
of the presence of the superconducting film
generating the proximity-induced pairing in the wire (or CNT),
one can change the chemical potential $\mu$ in the wire via a
gate voltage.
Our first step below is to show that experimentally available\cite{karma}
spiral magnetic fields offer strong effective SOI in CNTs, see also
Ref.~\onlinecite{braunecker0}.
As has been discussed in Ref.~\onlinecite{flensberg1},
it is possible to apply a spiral magnetic field to the CNT despite of the
presence of the superconductor.
We then proceed with a symmetry analysis of all possible
proximity-induced pairing terms in the low-energy theory
for a CNT in contact to an $s$-wave BCS superconductor.
Employing experimentally realistic parameters, we find generic gap closings
upon variation of a single control parameter,
usually corresponding to zero-energy states (``zero modes'') with finite
quasi-momentum $k$.  The gap closing can be probed experimentally
by tunneling spectroscopy.  For small-bandgap semiconducting CNTs
with magnetic field in the same plane, however, the zero mode is non-degenerate
and at $k=0$. This implies that Majorana bound states
(MBSs)\cite{hasan,qizhang,beenakker,sato,lutchyn,felix}
form at the tube ends. Majorana fermions may be useful for
topological quantum computation,\cite{nayak} and their
realization is now actively pursued in many different material
systems.\cite{beenakker}
Using InSb nanowires, a set-up similar to Fig.~\ref{fig1}, but
with intrinsically strong SOI instead of the spiral magnetic field, has
been studied experimentally, and clear evidence for the
predicted\cite{lutchyn,felix} MBSs at the ends of the nanowire was
reported from  tunnel spectroscopy.\cite{leo}
For quantum computation applications of MBS
states,\cite{nayak} CNTs could yield an attractive alternative
to InSb nanowires.  Apart from the unique electronic and mechanical properties
of CNTs and their wide availability,  MBS networks required for
braiding operations\cite{natphys,karsten1} could be
implemented by crossing CNTs.  Such CNT crossings have  been
experimentally realized already a decade ago.\cite{fuhrer,janssen}

MBSs in CNTs have also been proposed in two other papers,
relying on either the intrinsic curvature-induced SOI\cite{sau}
or\cite{loss1} on an electric-field induced SOI.\cite{loss}
Our paper is more closely related to the Sau-Tewari proposal,\cite{sau}
but differs in two important regards.  First, the spiral magnetic field
induces a much stronger SOI than the curvature-induced SOI in CNTs.
Second, this field also generates an effective periodic potential
along the CNT, which automatically breaks valley ($K,K'$) degeneracy
and implies a greatly enhanced robustness of the zero mode sector
from which we explicitly construct the MBS.  As  a consequence,
while the valley mixing parameter $\Delta_{KK'}$ is essential
in the Sau-Tewari proposal,\cite{sau} it plays no significant role
in the parameter space relevant for MBS generation in our proposal.
For completeness, we nonetheless keep $\Delta_{KK'}$ in our model Hamiltonian.

The structure of this article is as follows.
In Sec.~\ref{sec2}, we introduce the low-energy model for a CNT in
a spiral magnetic field.  We then continue in Sec.~\ref{sec3} with
a general analysis of the proximity-induced pairing terms appearing
in the Hamiltonian.  Results for the band structure are presented in
Sec.~\ref{sec4}, both for armchair tubes (Sec.~\ref{sec4a}) and for
small-bandgap semiconducting tubes (Sec.~\ref{sec4b}).  In the latter
case, in Sec.~\ref{sec4c} we explicitly construct the MBS wavefunction
when the spiral magnetic field is in the same plane as the CNT.
The role of electron-electron interactions is briefly addressed
in Sec.~\ref{sec4d}, and we conclude in Sec.~\ref{sec5}.

\section{CNT in a spiral magnetic field}
\label{sec2}

In the absence of a superconducting
substrate, the single-particle Hamiltonian for a straight CNT
along the $y$-axis reads\cite{cntreview}
\begin{eqnarray}\label{h0}
H_0 &=& -i\hbar v_F \partial_y \sigma_y  + E_g \sigma_x +
(e v_F R/2) B_y  \eta_z\sigma_x \\
\nonumber &+& \Delta_{KK'}\eta_x - \frac{g_e \mu_B}{2} {\bf B}(y)\cdot {\bf s}
\end{eqnarray}
with Pauli matrices $\sigma_{x,y,z}$ ($\eta_{x,y,z}$) in
sublattice (valley) space, where the two sublattices correspond
to the two carbon atoms forming the basis of the honeycomb lattice
and the two valleys denote the $K,K'$ points in the first
Brillouin zone;  Pauli matrices $s_{x,y,z}$ act in spin space.
This Hamiltonian acts on slowly varying Bloch envelope functions near
the $K,K'$ points, i.e., a state with quasi-momentum $k=0$
(with $-i\hbar \partial_y\to k$) sits right at the $K$ (or $K'$) point.
In Eq.~\eqref{h0} we omit the intrinsic
 SOI\cite{kuemmeth,jespersen,ando,egger,paco,weiss} since the spatially
dependent magnetic field ${\bf B}(y)$ will generate much larger couplings.
In the Zeeman term, the Bohr magneton is $\mu_B$ and we use $g_e=2$
for the Land{\'e} factor. Note that the orbital field along the CNT
($y$-)axis favors valley polarization in $z$-direction.
We also added the standard valley mixing term,\cite{kuemmeth,orb}
$\Delta_{KK'}$, which arises due to residual elastic disorder and
favors valley polarization in $x$-direction.

Writing the magnetic field in polar coordinates, cf.~Fig.~\ref{fig1},
\begin{equation}\label{field}
{\bf B} = B(y) \left (\begin{array}{c} \cos[\varphi(y)] \sin[\vartheta(y)]\\
\sin[\varphi(y)] \sin[\vartheta(y)] \\ \cos[\vartheta(y)] \end{array}\right),
\end{equation}
we next perform a unitary transformation,\cite{braunecker0,flensberg1}
\begin{equation}\label{unita}
U(y)=e^{\frac{i}{2} \vartheta [\sin(\varphi)  s_x - \cos(\varphi) s_y]},
\end{equation}
aligning the local spin quantization axis with the magnetic field direction.
The unitarily transformed Hamiltonian, $\tilde H_0=U^\dagger H_0 U$ with
Eq.~\eqref{h0}, then contains the effective SOI\cite{braunecker0}
\begin{eqnarray}\label{hprime}
\tilde H' &=&-i\hbar v_F \sigma_y U^\dagger
\partial_y U=  \frac{\hbar v_F}{2} \sigma_y {\bf a}(y) \cdot {\bf s}, \\
{\bf a}&=& \frac{d\vartheta}{dy}
\left(\begin{array}{c} \sin \varphi\\ -\cos\varphi\\ 0 \end{array}\right)
+\frac{d\varphi}{dy} \left(
\begin{array}{c} \cos\varphi\sin\vartheta\\ \sin\varphi \sin\vartheta
\\ 1-\cos \vartheta\end{array}\right) . \nonumber
\end{eqnarray}
Here we consider spiral magnetic field configurations, where
$\vartheta(y)=y/\lambda$ in Eq.~\eqref{field} and
both the field strength $B$ and the angle $\varphi$ are constant.
The case $\varphi=0$ (field in a plane perpendicular to CNT)
could, for instance, be realized using the hyperfine nuclear
fields discussed by Braunecker \textit{et al.},\cite{braunecker} and
the resulting SOI formally coincides with the
electric-field induced SOI studied in Ref.~\onlinecite{loss},
where we also recover their helical state solutions.
The case $\varphi=\pi/2$ (field in the same plane as the CNT) has
been realized experimentally using magnetic superlattices.\cite{karma}
Note that for arbitrary $\varphi$, the energy scale
$\delta = \hbar v_F/(2\lambda)$ associated to the ``pitch'' length $\lambda$
sets the effective SOI strength.  Using a typical value
$\lambda\approx 250$~nm, $\delta$ is several orders of
magnitude larger than the previously discussed spin-orbit couplings in CNTs.

When including proximity-induced pairing, it is convenient to work with Nambu
spinors, $\Psi^\dagger(y)=(\psi_\uparrow^\dagger,
\psi^\dagger_\downarrow,\psi_\downarrow,-\psi_\uparrow)$,
where $\tilde H=\int dy \Psi^\dagger {\cal H} \Psi/2$.
In a spiral magnetic field, using Eq.~\eqref{hprime} and
Pauli matrices $\tau_{x,y,z}$ in particle-hole space,
the CNT Hamiltonian reads
\begin{eqnarray}\nonumber
{\cal H} &=& \left[ -i\hbar v_F \sigma_y \partial_y
+ E_g \sigma_x  + A \sin(y/\lambda) \eta_z\sigma_x -\mu\right] \tau_z
\\  &-& \nonumber \mu_B B s_z + \delta \sigma_y [ \sin(\varphi) s_x -
\cos(\varphi) s_y ] \tau_z \\  &+& \Delta_{KK'} \eta_x \tau_z
+  {\cal H}_\Delta, \label{general}
\end{eqnarray}
where we added the chemical potential $\mu$ and a
proximity-induced pairing term  ${\cal H}_\Delta$.
Note that an orbital field in tube direction causes a periodic
potential with amplitude
\begin{equation}\label{adef}
A =\frac{ e v_F}{2} R B \sin \varphi.
\end{equation}
Experimentally, it turns out that this estimate for $A$
in some devices is enhanced by a factor $2$ to $3$ of unknown origin.\cite{orb}
Without  SOI and superconducting pairing,
periodic potentials have recently been studied theoretically in
CNTs\cite{levitov,novikov} and  graphene.\cite{steven}

At this point, we briefly comment on the 
time-reversal symmetry (TRS) properties of ${\cal H}$.  
TRS requires that ${\cal H}$ commutes with the
anti-unitary time-reversal operator ${\cal T}=is_y\eta_x {\cal C}$,
where ${\cal C}$ denotes complex conjugation and ${\cal T}^2=-1$.
Requiring TRS for ${\cal H}_\Delta$,
the only terms in Eq.~\eqref{general} violating TRS
are the Zeeman term $\propto B$ and the orbital flux $\propto A$.

\section{Proximity-induced superconductivity}
\label{sec3}

Next we discuss the proximity effect due to an $s$-wave singlet
superconducting substrate. To that end, we first write down all
$s$-wave singlet pairing terms in the CNT which are consistent with TRS.
With unity operator $\eta_0$ ($\sigma_0$) in valley (sublattice) space,
we obtain the ``zoology'' of allowed pairing terms,
\begin{eqnarray}\label{zoo}
{\cal H}_\Delta &=&\sum_{i=0,x} \Delta_i
(\cos\chi_i \ \eta_x +\sin\chi_i\ \eta_0) \sigma_i\tau_x  \\
&+& \nonumber
\Delta_2  (\cos\chi_2 \ \eta_x +\sin\chi_2\ \eta_0) \sigma_y\tau_y
\\ &+& \nonumber
\Delta_3 (\cos\chi_3 \ \sigma_x + \sin\chi_3 \ \sigma_0)\eta_y \tau_x +
\Delta_4 \eta_y\sigma_y\tau_y .
\end{eqnarray}
Terms $\propto \sigma_z$ ($\propto \eta_z$)
correspond to different substrate couplings for the two sublattices
(valleys) and have been omitted in Eq.~\eqref{zoo}.
In a generic situation, such asymmetries are extremely small\cite{smitha}
except for very thin CNTs, where the low-energy approach does not
apply in any case.  We also did not include TRS-invariant terms
$\propto s_{x,y,z}$ in Eq.~\eqref{zoo}, e.g., $s_x\sigma_y\tau_x$
or $s_z\tau_y$.  Such terms describe triplet pairing
in the CNT, which cannot be generated by coupling to
an $s$-wave superconductor.

The large number of parameters in Eq.~\eqref{zoo} can
now be greatly reduced by resorting to physical arguments.
The simplest and most likely dominant contribution to
the proximity effect comes from intra-sublattice (same orbital)
pairing: In a microscopic lattice model, ${\cal H}_\Delta=
\sum_i c_{i,\uparrow} c_{\downarrow,i} \Delta_i$, where
$i$ runs over all atoms in the honeycomb lattice.
This pairing mechanism corresponds to the three
terms $\propto \sigma_0$ in Eq.~\eqref{zoo}.
Another possibility comes from pairing between nearest-neighbor atoms
(different sublattices).
This mechanism explains all remaining terms [$\propto \sigma_{x,y}$] in
Eq.~\eqref{zoo}, but it has a much smaller amplitude because the
two sublattices are not commensurate. This implies that
the $2k_F$-oscillatory anomalous Green's function in the
superconductor basically averages out.
Neglecting these subleading contributions, we are left with the terms
$\propto \tau_x,\eta_x\tau_x$ and $\eta_y\tau_x$ in Eq.~\eqref{zoo}.
Next, note that the $\eta_x\tau_x$ and $\eta_y\tau_x$ terms,
which connect different valleys, are unitarily equivalent
in the absence of $KK'$ mixing.  Since the gap closings reported below
are also found for $\Delta_{KK'}=0$, we omit, say, the
$\eta_y\tau_x$ term in ${\cal H}_\Delta$.  We then arrive at
\begin{equation}\label{zoo1}
{\cal H}_\Delta = \Delta (\cos\chi \ \eta_x+\sin\chi\ \eta_0) \tau_x,
\end{equation}
with the proximity-induced gap $\Delta$ and
the ``pairing angle'' $\chi$. This angle interpolates between pure
inter-valley ($\chi=0$) and intra-valley ($\chi=\pi/2$) pairing.
The actual value for $\chi$ depends on how strongly rotational
symmetry around the CNT axis is broken by the presence of the substrate.
If rotational symmetry stays intact, different valleys form time-reversed
partner states and have to be paired,\cite{loss1,morpurgo}
resulting in $\chi=0$.  On the other hand,
TRS-invariant intra-valley pairing ($\chi=\pi/2$)
dominates for strongly broken rotational symmetry.\cite{sau}

Let us now briefly consider the case without the spiral magnetic field,
$A=B=\delta=0$ in Eq.~\eqref{general}, where  $\eta_x=\eta=\pm$ is
conserved.  Introducing the two proximity gap scales
\begin{equation}\label{deltaeta}
\Delta_{\eta}= \Delta \left|\sin\chi+\eta\cos\chi \right|,
\end{equation}
the dispersion relation follows from
\begin{equation}\label{mode2}
E^2_{\eta,\pm}(k)=\left(\eta\Delta_{KK'}-\mu\pm
\sqrt{(\hbar v_F k)^2+E_g^2} \right)^2+\Delta_\eta^2.
\end{equation}
When only one pairing term is present ($\chi=0$ or $\chi=\pi/2$),
the two gap scales coincide, $\Delta_\pm=\Delta$, and
the two pairing mechanisms cannot be distinguished unless the
magnetic field is also present.  Otherwise, however,
the pairing angle is detectable since then $\Delta_+\ne \Delta_-$.
In fact, the dispersion relation may become gapless for
$\eta=-$ and $\chi=\pi/4$.

\section{Results}\label{sec4}

Analytical diagonalization of the Hamiltonian
\eqref{general} with ${\cal H}_\Delta$ in Eq.~\eqref{zoo1} is not possible
except for special cases, and in general we have to resort to numerics.
Bloch's theorem implies that eigenstates for the $n$th
energy band, $E_n(k)$, are of the form
\begin{equation}\label{bloch}
\Psi_{k,n}(y) = e^{iky} \sum_{m\in\mathbb{Z}} \sum_{\nu} e^{imy/\lambda}
\Phi^{(n)}_{m,\nu}
\end{equation}
with the multi-index $\nu=(\sigma,\eta,s,\tau)$, where
$\sigma_z|\sigma=\pm\rangle=\sigma|\sigma\rangle$ (and so on).
The quasi-momentum $k$ is taken in the first Brillouin zone,
$-1\le 2k\lambda\le 1$, and $\Phi_{m,\nu}^{(n)}$ determines the
normalized eigenstate.
In this basis, all matrix elements of ${\cal H}$ in Eq.~\eqref{general}
except for the periodic potential $\propto A$ are diagonal in
$m$. The $A$ term couples $m$ and $m\pm 1$ states, and
diagonalization of the resulting Hamiltonian matrix yields
$E_n(k)$ and the eigenstates (\ref{bloch}).
We always find $E_n(-k)=E_n(k)$ and thus show only half of the Brillouin zone
below.

The model parameters in Eqs.~\eqref{general} and \eqref{zoo1} are chosen as
follows. We take an effective SOI scale $\delta=2$~meV corresponding
to magnetic pitch length $\lambda\approx 250$~nm. This
is a typical value for magnetic superlattices, see Ref.~\onlinecite{karma}.
The Zeeman scale is $\mu_B B=0.5$~meV (for $B\approx 5$~T), and
taking into account Ref.~\onlinecite{orb}, the orbital field implies
the periodic potential amplitude
$A=A_0\sin\varphi$, see  Eq.~\eqref{adef}, with $A_0=2$~meV
(for $R\approx 1$~nm).
For the proximity-induced gap, $\Delta=0.3$~meV is appropriate for
Nb substrates.\cite{grove}  The $KK'$ mixing scale is taken as
$\Delta_{KK'}=0.2$~meV,\cite{jespersen,orb}
which is a typical order-of-magnitude value for this phenomenological
parameter.  We note again that in contrast to the
proposal of Ref.~\onlinecite{sau},  $\Delta_{KK'}$
is not necessary for the MBS generation reported in Sec.~\ref{sec4d}.
Next, for the bandgap we choose either
 $E_g=0$ (armchair case) or $E_g=3$~meV (small-bandgap semiconducting CNT).
For the field angle $\varphi$, we take mostly $\varphi=0$ or
$\varphi=\pi/2$, but we have checked that small deviations from these
values do not cause qualitative changes (see also below for a more
detailed discussion).  This leaves us essentially with two free
parameters, namely the chemical potential $\mu$ and the pairing angle $\chi$.
(The pairing angle, however, is hard to change in an actual experiment.)

\subsection{Armchair CNTs}
\label{sec4a}

We begin with the armchair case, $E_g=0$,
where $\eta_x=\eta=\pm$ is conserved.
For field angle $\varphi=0$, the spectrum for
Eq.~\eqref{general} with \eqref{zoo1} can be obtained analytically
by squaring ${\cal H}$ twice.  Noting that also
$\sigma_y=\sigma=\pm$ is conserved,
\begin{eqnarray}\label{noarm}
E^2_{\eta,\sigma,\pm}(k) &=& \epsilon_{k\eta\sigma}^2 + (\mu_B B)^2 +
\delta^2 +\Delta_\eta^2 \\ \nonumber
&\pm& 2 \sqrt{\epsilon_{k\eta\sigma}^2
[\delta^2+(\mu_BB)^2] + (\mu_B B \Delta_\eta)^2}
\end{eqnarray}
with $\epsilon_{k\eta\sigma}=\sigma v_F k-\mu+\eta\Delta_{KK'}$ and
$\Delta_\eta$ in Eq.~\eqref{deltaeta}.  Equation \eqref{noarm} does
not permit zero-energy solutions for finite $\delta$ and $\Delta_\eta$,
and there is no gap closing for this configuration.

\begin{figure}[t]
\centering
\includegraphics[width=8cm]{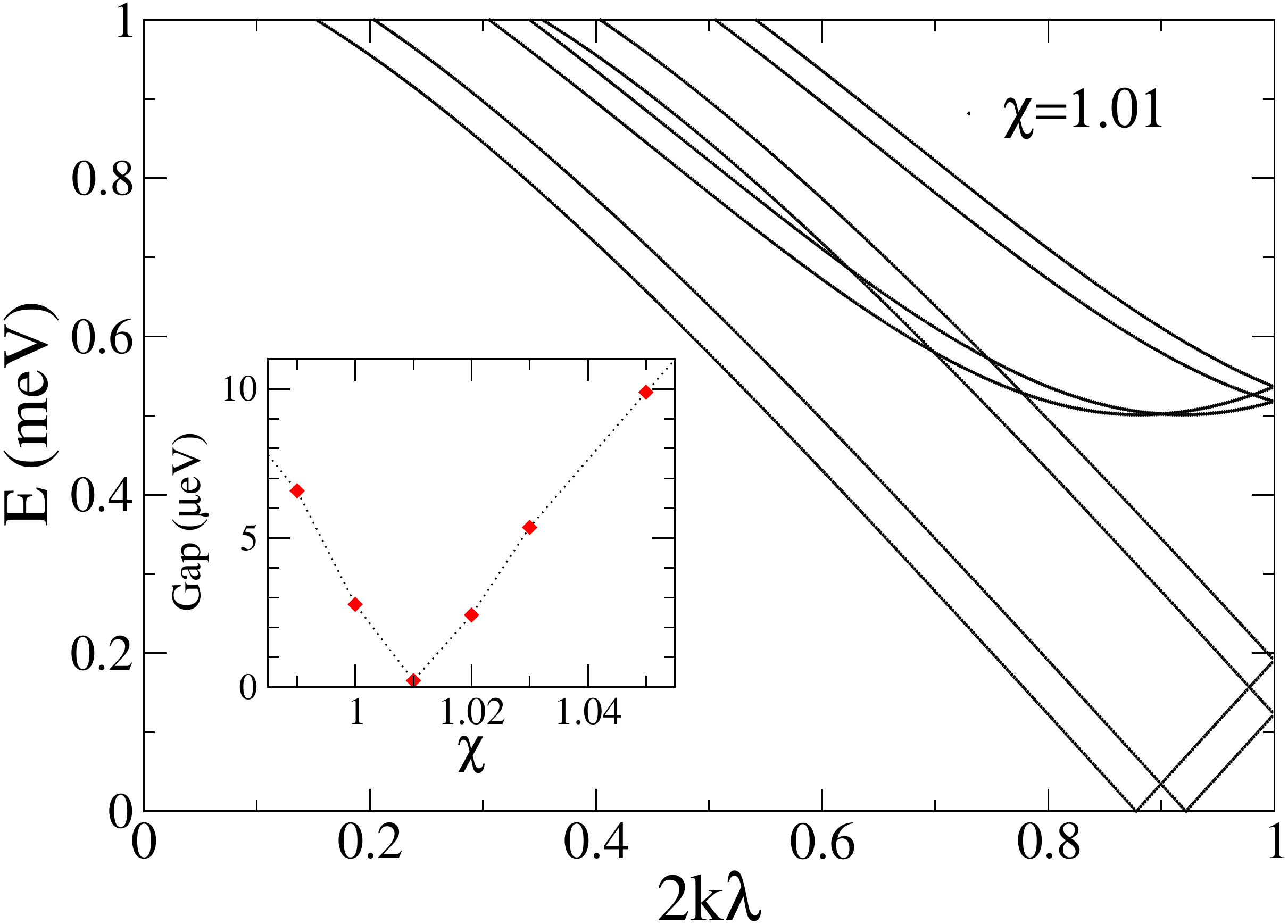}
\caption{\label{fig2} (Color online)  Dispersion relation for the
armchair case with field angle $\varphi=\pi/2$, chemical potential
$\mu=0$, and pairing angle $\chi=0.32\pi=1.01$.
Note that there are two positive $k$ where the gap vanishes, plus
the respective $k<0$ states with $E_n(-k)=E_n(k)$, where we show only
half of the first Brillouin zone.
Inset: Gap closing as a function of $\chi$. Red diamonds give numerical
results, the dotted line is a guide to the eye only.
} \end{figure}

Turning to $\varphi=\pi/2$, numerical solution at chemical potential
$\mu=0$ yields the energy bands in Fig.~\ref{fig2}. Notably, the gap
closes for $\chi\simeq 1.01$, see inset of Fig.~\ref{fig2}, with
two pairs of zero modes at finite momenta $k$.  Note that for both
values of $k$, each zero mode is still twofold degenerate because of
its partner state at $-k$. The linear (massless Dirac fermion) dispersion is
clearly visible in Fig.~\ref{fig2}.
Allowing also for a finite but small value of $\mu$, we find that only
one of the gap closings seen in Fig.~\ref{fig2} persists.

\subsection{Small-bandgap semiconducting CNTs}
\label{sec4b}

Next we study the case of a small bandgap; for concreteness,
$E_g=3$~meV. For field angle $\varphi=0$, we find just a single zero mode
at finite $k$, along with its partner state at $-k$.
To give concrete numbers, a finite-$k$ zero mode
was found for $\mu=2.96$~meV and $\chi=0.76$.
This behavior is qualitatively similar to the armchair case
with $\varphi=\pi/2$ and finite $\mu$.

\begin{figure}[t]
\centering
\includegraphics[width=8cm]{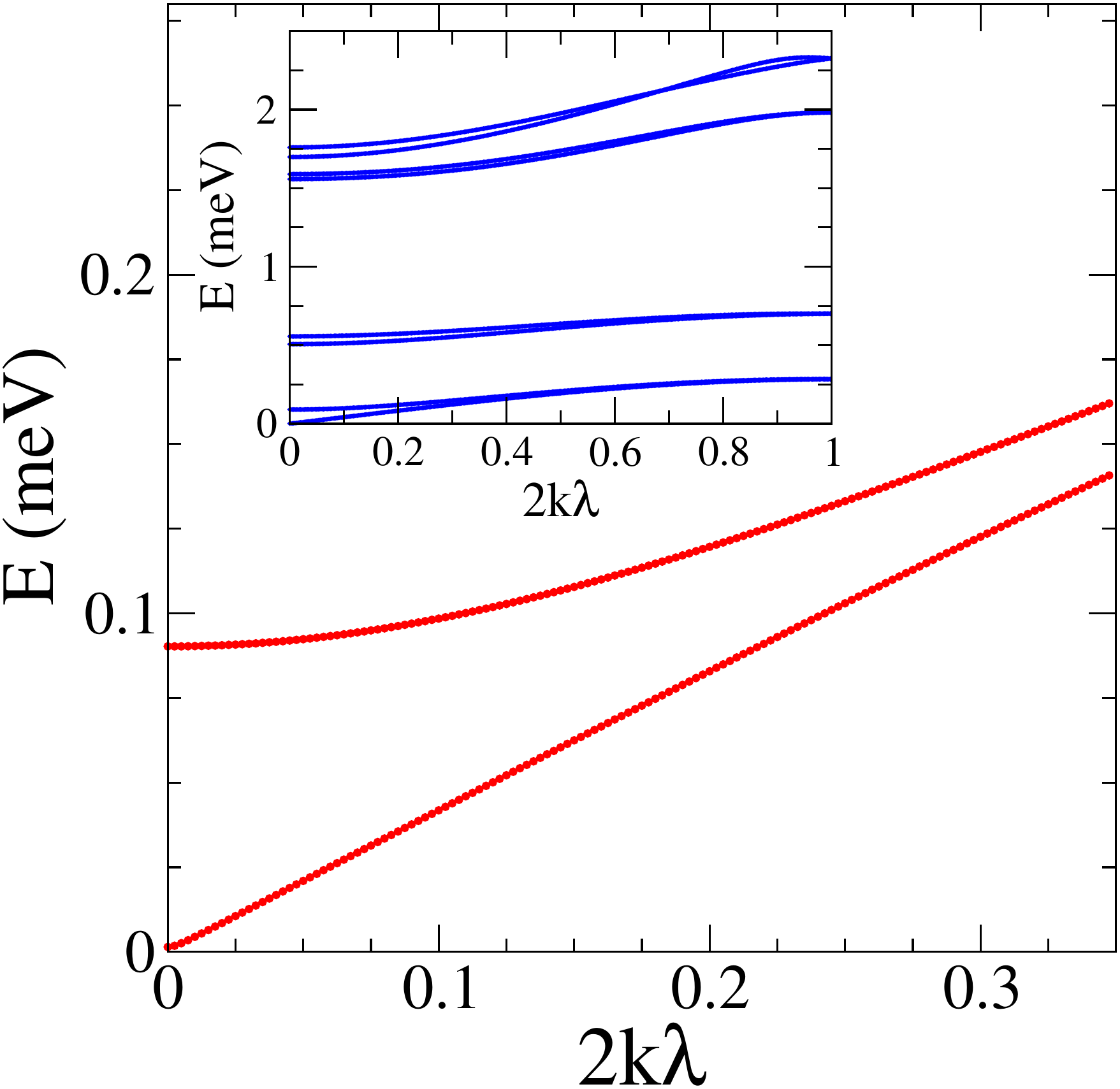}
\caption{\label{fig3} (Color online) Low-energy dispersion relation for
a CNT with bandgap $E_g=3$~meV, field angle $\varphi=\pi/2$, chemical potential
$\mu=2.86$~meV, and pairing angle $\chi=0.14$.
Note the linear dispersion near $k=0$, where the velocity is
$v\simeq 0.15 v_F$.  The inset shows the dispersion also for higher energies.
} \end{figure}

For a small-bandgap CNT and field angle $\varphi=\pi/2$, we
encounter a remarkably different situation with only a single
non-degenerate zero mode at $k=0$, see Fig.~\ref{fig3}.
Only in such a non-degenerate case,
single MBS formation is possible.\cite{beenakker}
Note that TRS has been broken by the applied magnetic field here;
otherwise the MBS must have an overlapping 
time-reversed partner.\cite{Fulga,Ryu}
This gapless state can be reached upon variation of $\chi$ through a
``magic angle,'' cf.~inset of Fig.~\ref{fig2}.  An experimentally
easier route is to change, for fixed $\chi$, the chemical
potential to the critical value through variation of a gate voltage.
We stress that this zero-energy state can be reached by tuning a
single parameter.

\begin{figure}[t]
\centering
\includegraphics[width=8cm]{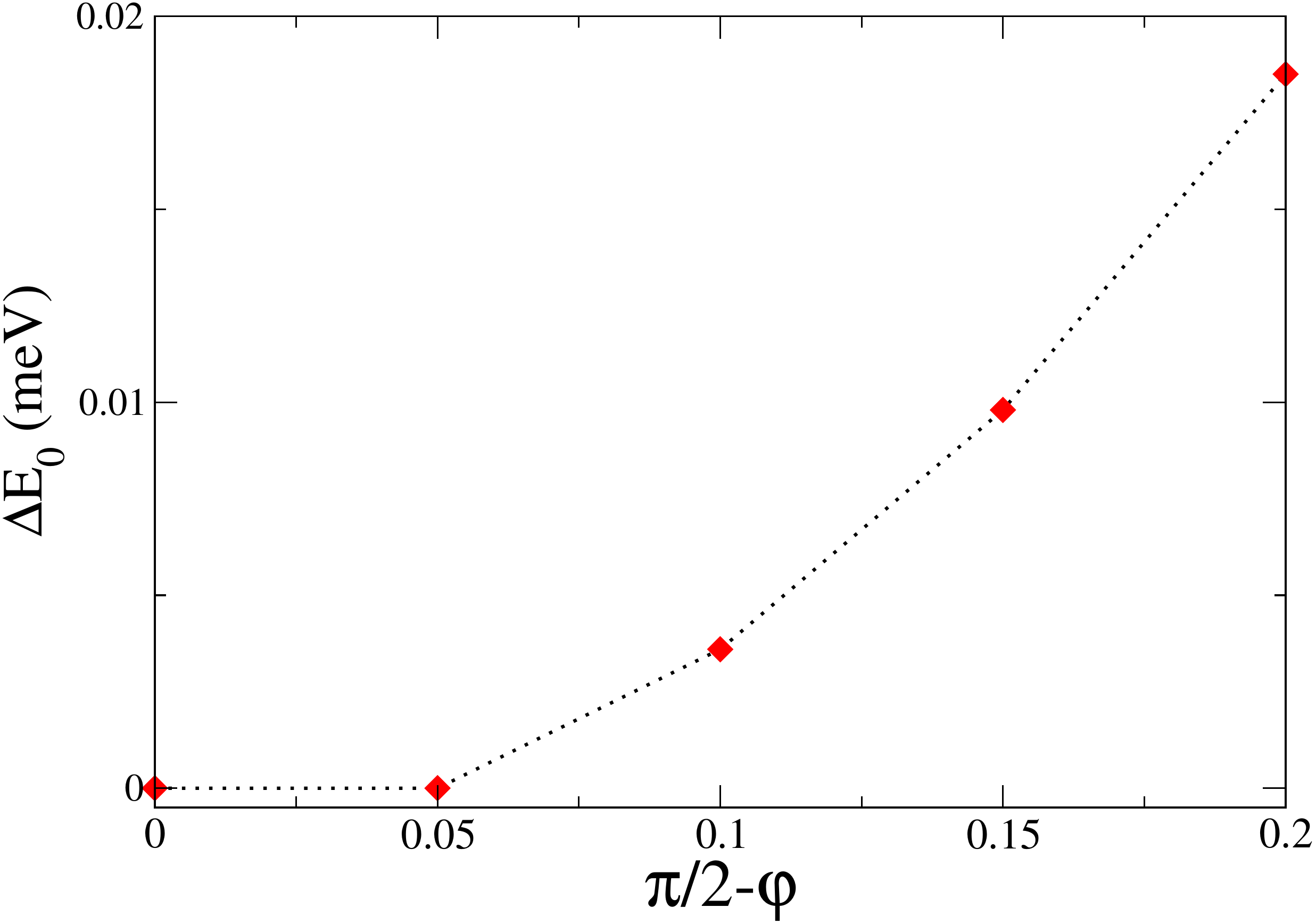}
\caption{\label{fig4} (Color online)  Band gap $\Delta E_0$ (at $k=0$)
for the parameters in Fig.~\ref{fig3} but with varying field
angle $\varphi$.  Red diamonds indicate numerical results, the dotted
curve is a guide to the eyes only.
} \end{figure}

We have also checked that the $k=0$ gap closing observed in
Fig.~\ref{fig3} stays robust against small changes of $\varphi$ (or
other model parameters).  To illustrate this point, we show the $k=0$ gap
$\Delta E_0$ as a function of $\varphi$ in Fig.~\ref{fig4}.
Clearly, for $\varphi$ near $\pi/2$, we find $\Delta E_0=0$, but
sufficiently large deviations will destroy the zero mode.

\subsection{MBS construction}
\label{sec4c}

For small-bandgap semiconducting
CNTs with $\varphi=\pi/2$, the $k=0$ zero mode in Fig.~\ref{fig3}
and the gap closing and reopening as $\mu$ is varied through its critical
value $\mu_c$ suggest that a MBS\cite{hasan,qizhang,beenakker} exists at
the interface of regions with $\mu<\mu_c$ and
$\mu>\mu_c$.  In practice, for $\mu<\mu_c$, a MBS should then form at
the CNT ends.  To model this situation, let us consider
$\mu(y)=  \mu_c - \alpha y$ with $\alpha>0$, where a
MBS is expected near $y=0$.  We explicitly construct the MBS
wavefunction by first projecting the full Hamiltonian \eqref{general}
to the Hilbert space spanned by the massless Dirac fermions,
${\cal H}\Psi_k^{(\pm)}= \pm vk\Psi_k^{(\pm)}$, with
$v\simeq 0.15v_F$ from Fig.~\ref{fig3} and $\Psi_k^{(\pm)}(y)$
known numerically; the eigenvectors $\Phi^{(\pm)}_{m,\nu}$ in
Eq.~\eqref{bloch} are evaluated at $k=0$.
With Pauli matrices $\tilde \tau_{x,y,z}$ acting in the
Hilbert space spanned by $\Phi^{(+)}$ and $\Phi^{(-)}$,
the projected low-energy Hamiltonian is
\begin{equation}\label{mbs}
{\cal H}_p=-i\hbar v \tilde \tau_z \partial_y +\alpha y\tilde \tau_x.
\end{equation}
The second term is due to the spatial variation of the chemical potential,
where we find $\tau_z\simeq \tilde \tau_x$ in the zero-mode basis by using
the numerically obtained  eigenvectors $\Phi^{(\pm)}$.
Writing $\Phi=c_+\Phi^{(+)}+c_-\Phi^{(-)}$ (where $c_\pm$ are
complex numbers),  the state
\begin{equation}
\Psi(y)\propto e^{-\alpha y^2/2\hbar v} \Phi,\quad \tilde \tau_y\Phi=-\Phi,
\end{equation}
then yields a zero-energy Majorana fermion solution, ${\cal H}_p\Psi=0$,
localized near $y=0$. This state is topologically protected by the
gap to the next excited state. For the parameters of Fig.~\ref{fig3},
this ensures MBS robustness for temperatures $T\alt 1$~K.

\subsection{Electron-electron interactions}\label{sec4d}

So far we have ignored electron-electron interaction effects beyond
mean-field theory.  Following the well-known fact
that interactions destabilize the Fermi liquid phase in one dimension,
it has been suggested,\cite{egger1,kane1} and
subsequently observed,\cite{exp1,exp2,exp3} that
CNTs display Luttinger liquid behavior. The Luttinger liquid
phase is a strongly correlated phase, and therefore one should
be careful in applying noninteracting theories to CNTs as done here.
Importantly, the experiments in Refs.~\onlinecite{exp1,exp2,exp3}
were performed without close-by metallic gates such
that the long-range character of the Coulomb interaction was important.
However, scanning tunneling spectroscopy for CNTs
deposited directly on a metallic substrate did not show pronounced
interaction phenomena (for a review, see Ref.~\onlinecite{sts}),
presumably due to the strong screening of the
Coulomb potential by the substrate.
Similarly, we expect that the presence of the
superconducting substrate, see Fig.~\ref{fig1}, drastically reduces
the effective interaction strength. The remaining weak interactions,
on the other hand, are then not expected to destroy the MBS state or the
emerging Dirac fermions discussed above.  This topological stability
against weak interactions has been discussed in detail in several
recent works,\cite{int1,int2,int3} and we refrain from repeating their
analysis here.

\section{Concluding remarks}
\label{sec5}

In this work, we have studied the low-energy
bandstructure of CNTs with effective SOI (due to a spiral magnetic field)
and proximity-induced pairing. As discussed in Sec.~\ref{sec4d},
in the presence of the superconducting
substrate, interactions are screened and our single-particle
approach should be useful. Despite of the combined presence of a bandgap,
the strong SOI, the orbital periodic potential, and the proximity gap,
we find zero modes with associated massless Dirac fermions.
The gap closings should guide future tunneling spectroscopy experiments:
For parameters near a gap closing condition, topological end
states will appear and can be observed as peaks in the $dI/dV$
curve.\cite{beenakker}  For finite-$k$ zero modes, this corresponds to
degenerate Majorana states at each end, while for the $k=0$ gap closing,
a single localized Majorana mode results.

\acknowledgments

This work was supported by the SFB-TR 12 of the DFG (R.E.) and by
The Danish Council for Independent Research $|$ Natural Sciences (K.F.).

\end{document}